# An annotated list of bivalent chromatin regions in human ES cells: a new tool for cancer epigenetic research

**Franck Court[1,2,3], Philippe Arnaud[1,2,3]**

[1]CNRS-UMR 6293, Clermont-Ferrand, 63001, France

[2]INSERM-U1103, Clermont-Ferrand, 63001, France

[3]Université Clermont Auvergne, GReD Laboratory, Clermont-Ferrand, 63000, France

**Correspondence to:** Philippe Arnaud, **email:** philippe.arnaud@udamail.fr

Franck Court, **email:** franck.court1@udamail.fr

## ABSTRACT

CpG islands (CGI) marked by bivalent chromatin in stem cells are believed to be more prone to aberrant DNA methylation in tumor cells. The robustness and genome-wide extent of this instructive program in different cancer types remain to be determined. To address this issue we developed a user-friendly approach to integrate the stem cell chromatin signature in customized DNA methylation analyses. We used publicly available ChIP-sequencing datasets of several human embryonic stem cell (hESC) lines to determine the extent of bivalent chromatin genome-wide. We then created annotated lists of high-confidence bivalent, H3K4me3-only and H3K27me3-only chromatin regions. The main features of bivalent regions included localization in CGI/promoters, depletion in retroelements and enrichment in specific histone modifications, including the poorly characterized H3K23me2 mark. Moreover, bivalent promoters could be classified in three clusters based on PRC2 and PolII complexes occupancy. Genes with bivalent promoters of the PRC2-defined cluster displayed the lowest expression upon differentiation. As proof-of-concept, we assessed the DNA methylation pattern of eight types of tumors and confirmed that aberrant cancer-associated DNA hypermethylation preferentially targets CGI characterized by bivalent chromatin in hESCs. We also found that such aberrant DNA hypermethylation affected particularly bivalent CGI/promoters associated with genes that tend to remain repressed upon differentiation. Strikingly, bivalent CGI were the most affected by aberrant DNA hypermethylation in both CpG Island Methylator Phenotype-positive (CIMP+) and CIMP-negative tumors, suggesting that, besides transcriptional silencing in the pre-tumorigenic cells, the bivalent chromatin signature in hESCs is a key determinant of the instructive program for aberrant DNA methylation.

## INTRODUCTION

The proper development of higher organisms is a tightly regulated process in which epigenetic pathways are key determinants. Consistently, in addition to genetic lesions, epigenetic alterations are important actors in various human pathologies, including carcinogenesis. Specifically, aberrant DNA hypermethylation at gene promoter-associated CpG Islands (CGI/promoter) is a well characterized feature of cancer cells. In patients with glioma, colorectal or lung cancer, the CGI methylation pattern at some genes, such as *MGMT* [1], *SEPT9* [2] or *SHOX2* [3], can be used as a biomarker for diagnosis, prognosis or prediction of the response to therapies. Moreover, a CpG island methylator phenotype (CIMP), characterized by the concomitant hypermethylation of multiple CGIs, has been described in a subset of different tumor types [for review 4], including colorectal, gastric, breast and lung cancer as well as glioblastoma and hematological malignancies. Strikingly, CIMP-positive tumors exhibit specific molecular features, clinical prognosis and outcome compared with CIMP-negative tumors [4], thus underlying the potential of these signatures as cancer biomarkers.

Studies on the molecular bases of CGI hypermethylation suggest the existence of an instructive program that relies on the CGI chromatin signature in stem cells. Specifically, both candidate-based and genome-wide analyses highlighted that genes marked in embryonic or adult stem cells by repressive polycomb group proteins (PcG) and more specifically by a bivalent chromatin signature are prone to be aberrantly methylated in cancer cells [5–8]. CGIs/Promoters with bivalent chromatin are concomitantly marked by the 'active' H3K4me3 and the 'repressive' H3K27me3 marks. Bivalent chromatin domains are thought to repress gene transcription through H3K27me3, while keeping genes 'poised' for alternative fates induced by specific developmental cues upon stem cell differentiation [9, 10]. Consistently, overlaps between the gene expression signatures of stem cells and aggressive tumors have been reported [11]. Altogether, these observations led to the hypothesis that besides mediating the stable silencing of tumor suppressor genes, the main consequence of CGI hypermethylation is to aberrantly maintain cancer cells in a "plastic" stem-cell like state (poor differentiation capacity and unlimited self-renewal) that contributes to cancer initiation and progression [5, 12, 13]. However, the observation that CIMP-positive tumors are associated with a better clinical prognosis [14, 15] challenged this hypothesis. Also, recent integrative genome-wide analyses revealed that aberrant CGI hypermethylation affects primarily genes that are already repressed in the matched normal tissue, indicating that most of the aberrantly methylated genes are not involved in carcinogenesis [8, 16, 17]. To account for these observations, Sproul and Meehan proposed an alternative hypothesis in which the stable repression brought by aberrant DNA hypermethylation at CGIs/promoters restricts the epigenetic plasticity of cancer cells and their ability to adapt following environmental changes, such as during metastasis formation or treatment, thus acting as a protective mechanism against cancer progression [18].

Evaluating if these two possibilities co-exist in cancer cells, with respect to cancer subtypes and/or stages of the disease, emerge thus as an important issue to formally characterize the role of DNA hypermethylation in tumors cells. For this, we need to precisely determine genome-wide to which extent the stem cell chromatin signature pre-marks CGIs for hypermethylation in cancer cells. DNA methylome studies in a variety of cancer cells have benefitted from the development of cost-effective and normalized tools, such as the Infinium HumanMethylation450 (HM450K) BeadChip Arrays or the recently released Infinium MethylationEPIC Arrays (both from Illumina) that allow comparing results between laboratories. However, the analysis of chromatin signatures in stem cells is hampered by the bioinformatics skills needed to handle the publicly available genome-wide ChIP-sequencing datasets.

Alternatively, already processed lists of chromatin signatures could highly facilitate this kind of approach. Three initial studies have provided a list of the bivalent chromatin regions in human embryonic stem cells (hESCs) [19–21]. In these three studies, data generated following genome-wide chromatin immunoprecipitation using antibodies against H3K4me3 or H3K27me3 were merged to identify genes carrying both modifications. Although all three studies limited their analysis to the regions surrounding the transcription start site (TSS), the number of identified bivalent promoters greatly varies, from 1798 in [20], 2500 in [21] and more than 5500 in [19]. Moreover, these three lists show an incomplete and limited overlap [22]. This discrepancy can be explained by the different experimental approaches (*i.e.*, sequencing *vs* microarray), insufficient sequencing depth [23] and the use of relaxed statistical criteria. It could also reflect the variability between hESC lines, because each study used a different line: H1 [19], hES3 [20] an H9 [21]. A more recent study precisely improved the accuracy and characterization of bivalent regions in both mouse and human genomes by relying on datasets for several independent ESC lines [24]. This study provided the chromatin signature of all annotated promoters in hESCs and identified 4979 of them as being bivalent [24].

However, none of these four studies gave the exact genomic coordinates of the bivalent regions nor investigated the presence of a bivalent signature genome-wide. In addition, the spreadsheet-based files used to present the results are not easy to handle for integrative analyses using DNA methylation array data.

These observations stress the need to establish an annotated list of high-confidence (HC) bivalent regions in hESCs that can be easily used by researchers and clinicians to investigate the relationship between the bivalent chromatin signature in hESCs and methylation defects in cancer cells. To this aim, we used publicly available ChIP-sequencing datasets of several hESC lines to determine and characterize bivalent chromatin regions genome-wide. We created annotated lists of HC bivalent, H3K4me3-only and H3K27me3-only genomic regions and developed a user-friendly approach to integrate these chromatin signature features in DNA methylation analyses.

## RESULTS

### Creation of a list of HC bivalent genomic regions in hESCs

To establish an exhaustive list of the genomic regions marked by H3K4me3, H3K27me3 or both (*i.e.*, bivalent) in hESCs, we took advantage of publicly available ChIP-sequencing (ChIP-seq) datasets. In total, we collected 11 dataset series (*i.e.*, input; H3K4me3 and H3K27me3) each produced by the same laboratory and using the same hESC sample (Supplementary Table S1). By combining the dataset series for the same hESC line, we obtained pairs of input, H3K4me3 and H3K27me3 ChIP-seq datasets for five hESC lines: HUES48, HUES64, HUES6, I3 and H1 (Supplementary Table S2). As the sequencing depth is critical for obtaining statistically significant results [23], we ensured that each ChIP-seq dataset included at least 20 million uniquely mapped reads (Supplementary Table S2). For each hESC line, peaks were called in the H3K4me3 and H3K27me3 datasets using Macs1.4.2. Genome-wide, we identified between 26 354 and 50 312 peaks for H3K4me3 and between 56 163 and 92 099 for H3K27me3 (Figure 1A and Supplementary Table S3).

We next isolated regions enriched for both histone marks and defined as bivalent domains all regions where H3K4me3 and H3K27me3 peaks overlapped for at least 1Kbp (Figure 1B). The number of bivalent domains ranged from 7 756 in I3 cells to 10 266 in HUES6 cells (Figure 1C, Supplementary Table S3), giving a merged list of 12 402 bivalent domains. This variation between hESC lines could be explained by different culture conditions or antibodies used for ChIP and also the sequencing depth across samples (Supplementary Table S2). Indeed, the number of uniquely mapped reads correlated with the number of identified peaks (for H3K27me3, Pearson's correlation (r) = 0.86, α < 0.05; for H3K4me3 (r) = 0.85, α < 0.05).

A subset of the identified regions was hESC line-specific, particularly in HUES6 cells (10% of all identified bivalent domains) (Figure 1C, Supplementary Table S3). These cell line-specific regions tended to have a weaker enrichment for H3K27me3 and, to a lesser extent, for H3K4me3 compared with bivalent regions present in several cell lines (Figure 1D). By restricting the merged list to regions present in all five hESC lines, we obtained a list of 5 766 HC bivalent domains present in the hESC genome.

By using the same criteria (*i.e.*, more than 1Kbp in size and common to the five hESC lines), we also established a list of HC H3K4me3-only (*n* = 11 966) and H3K27me3-only (*n* = 16 361) regions (Supplementary Figure S1). These regions had a median size of about 2.5Kbp and generally did not exceed 3.5Kbp. H3K27me3-only regions were larger with a size up to 8.5kbp (Supplementary Figure S2A).

The full list of these regions, with their genomic coordinates, associated gene and main features, is available in Supplementary File S1. In addition, to facilitate their customized visualization in their genomic context, we produced a file ready for uploading on the UCSC genome browser (https://genome.ucsc.edu/) (Supplementary File S2).

## Most bivalent regions are localized in CGI-rich promoter

At the genomic level, CGIs clearly discriminated H3K27me3-only from H3K4me3-only and bivalent regions (Figure 1E, Supplementary Table S4). Specifically, 71% and 61% of bivalent and H3K4me3-only regions, but only 1.5% for H3K27me3-only regions overlapped with a promoter-associated CGI (CGI/promoter) region. This indicates that most CGI/promoters are not marked by H3K27me3 alone.

By including also regions not associated with CGIs, 74% of H3K4me3-only and 75% of bivalent regions were within a promoter. Conversely, only 24.5% of H3K27me3-only regions were associated with a promoter and most of them lacked CGIs (Figure 1E, Supplementary Figure S2B, Supplementary Table S4).

To determine the gene ontology term enrichment of each region type, we filtered for genes with several promoters and retained only those with a similar chromatin signature at all their marked promoters (Supplementary Figure S3). Genes with bivalent promoter regions were strongly enriched for cell differentiation pathways and development processes. On the other hand, basal metabolic processes, cell cycle and repair pathways were the main terms associated with genes with H3K4me3-only promoters (Figure 1F and Supplementary File S3).

Our strategy, based on an unbiased genome-wide approach also revealed that 26% and 25% of H3K4me3-only and bivalent regions, respectively, were in intergenic and gene body areas (Supplementary Table S4). This might reflect the presence of not yet annotated promoter regions. Sub-class of enhancers could also account for this enrichment. Indeed, using the enhancer lists defined by the Fantom5 project we found that 39% (563/1436) of the bivalent domains and 30% (1054/3502) of the H3K4me3-only regions not associated with a promoter were in regions defined as enhancers (Supplementary File S1).

## Molecular signature associated to bivalent promoter regions

We next aimed to characterize the genomic composition and molecular signature of the promoter-associated regions for each of these 3 categories. It has been proposed that CGI/promoters prone to be hypermethylated in cancer cells are depleted in retroelements at the transcriptional start site (TSS) [25]. Overall, we found that the three major classes of retrotransposons (LINEs, SINEs and long terminal repeats (LTRs)) were depleted at TSS; however, bivalent promoter regions showed the greatest depletion, particularly in the genomic region directly bordering the TSS (Figure 2A, Supplementary Figure S2C).

Using publicly available ChIP-seq data for the H1 and H9 hESC lines, we investigated the deposition of 22

histone modifications (Supplementary Table S5). Some marks known to be associated with active promoters or enhancer regions, such as H3K27ac and H3K79me1, were more specifically found at H3K4me3-only regions. Conversely, H3K4me2, H2AZ or H3K9ac marked also bivalent regions (Figure 2B, Supplementary Figure S2D). However, while H3K4me2 and H2AZ signals intensity was comparable at H3K4me3-only and bivalent promoters, H3K9ac signal was stronger at H3K4me3-only promoters. The poorly characterized H3K23me2 mark was present both at H3K4me3-only and bivalent promoters, but its mean signal intensity was stronger at bivalent promoters. H3K27me3-only regions were depleted for most of the 22 analyzed modifications, if not all. These findings suggest that some "active" marks are strictly enriched at active promoters, while others, similarly to H3K4me3, can co-exist with the repressive H3K27me3 mark in a poised configuration. This hypothesis is also supported by the observation that only the H3K4me3-only promoters (blue in the left histogram of Figure 2C) were associated with highly expressed genes in hESCs.

Besides histone modifications, the analysis of ChIP-seq data for 59 transcription factors in the hES1 cell line stressed that differently from H3K4me3-only promoters, transcription factor occupancy was low at H3K27me3-only promoters in hESCs (Supplementary Figure S4, Supplementary Table S6). The transcription factor occupancy at bivalent promoters showed shared and also specific features, compared with H3K4me3-only promoters, in agreement with their intermediate status (Figure 2D). As expected, components of the PRC2 complex that mediates H3K27me3 deposition were enriched at bivalent promoters (EZH2: 92.8% of bivalent *vs* 17.2% of H3K4me3-only promoters: Chi-squared test $p < 2.2^{e-16}$; SUZ12: 59.4% of bivalent *vs* 12.5% of H3K4me3-only promoters: Chi-squared test $p < 2.2^{e-16}$). Both promoter categories were also widely marked by the Polymerase II complex (PolII, P300, TBP), including the PolII form phosphorylated at Ser5 that correlates with transcriptional initiation [26]. Conversely, the general transcription factor IIF (GTFIIF1) that promotes transcriptional elongation [27] was mainly restricted to H3K4me3-only promoters (25.7% of bivalent *vs* 65.5% of H3K4me3-only: Chi-squared test $p < 2.2^{e-16}$), suggesting that at most bivalent promoters, the loaded PolII complex is in a unproductive pre-initiation state.

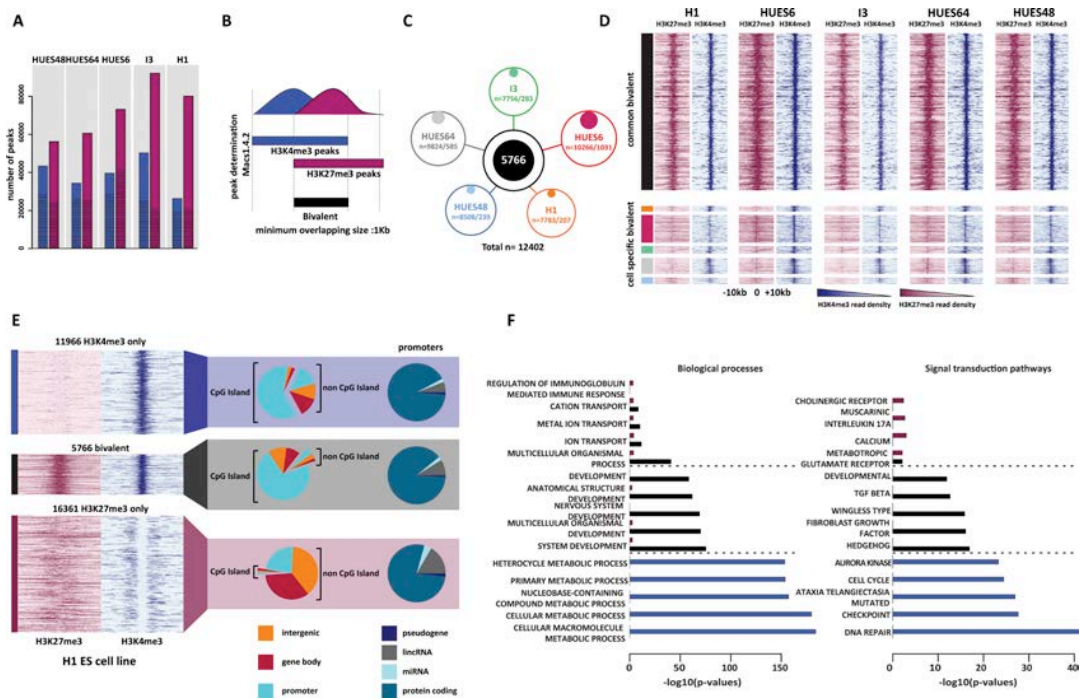

**Figure 1: Genome-wide identification of high-confidence bivalent domains in human ES cells.** (**A**) Number of H3K4me3 (blue columns) and H3K27me3 (purple columns) peaks in each hESC line. The proportion of peaks wider than 1Kbp is indicated by the hatched area within each column. (**B**) Schematic overview of the strategy used to identify bivalent domains. (**C**) Number of bivalent domains identified in each cell line (total/cell line-specific). In total (merged results from the five hESC lines) 12 402 bivalent domains were identified of which 5 766 were present in all five cell lines and were considered as high-confidence bivalent domains. (**D**) ChIP-seq read density of H3K27me3 (purple) and H3K4me3 (blue) for each hESC line in a ± 10Kbp window centered on the peak signal. Upper panel: high-confidence (*i.e.,* common to all five lines) bivalents domains. Lower panel; cell type-specific bivalent domains. Color code as in 1C). (**E**) Genomic features associated with high-confidence H3K4me3-only, bivalent and H3K27me3-only regions. ChIP-seq read density (H3K27me3: purple; H3K4me3: blue) is shown for the H1 cell line (see Supplementary Figure S1 for the other hESC lines). (**F**) Gene ontology terms enriched in H3K27me3-only (purple columns), bivalent (black columns) and H3K4me3-only (blue columns) promoters. For each category, the five highest terms are shown. The full list is available in Supplementary File S3.

## PRC2 and PolII complex occupancy defines clusters at bivalent promoters

As PRC2 and PolII complex occupancy can vary among bivalent promoters [21, 24], to further characterize the signature of bivalent promoters in hESCs, we investigated the signal density of EZH2 (a PRC2 component) and PolII, as well as of a dozen of other selected factors among which the TATA-box binding protein-associated factor1 (TAF1) and the Transcription Factor 12 (TCF-12; also named HEB), a component of nodal signaling in hESCs, were discriminating. Indeed, based on the signal density of these four factors, we could classify HC bivalent promoters in three clusters (Figure 3). Moreover, we observed a similar clustering also when we considered all HC bivalent regions, regardless of their promoter status (Supplementary Figure S5).

The first cluster included about 17% of bivalent promoters (748/4330) and was characterized by a sharp and strong signal for PolII and TAF1. Conversely, the second cluster (22% of bivalent promoters; 952/4330) displayed a strong signal for EZH2, and to a lesser extent, for TCF12, whereas PolII and TAF1 were depleted. The third and main cluster (60% of bivalent promoters) included promoters that were similarly marked by all these factors. Consistently, a marked enrichment for H3K27me3 and depletion for H3K9ac characterized bivalent regions from cluster 2 (Supplementary Figure S5). Functional gene enrichment analysis indicated that promoters belonging to cluster 2 controlled genes enriched in transcription factor activity ($p$ value $< 10^{-109}$) and regulation of transcription ($p$ value $< 10^{-62}$). Cluster 3 genes were associated with plasma membrane ($p$ value $< 10^{-42}$). In agreement with their molecular signature, genes controlled by bivalent promoters of cluster 1 showed the highest expression in hESCs. Conversely, expression of cluster 2 genes was low, similarly to that of genes controlled by H3K27me3-only promoters (Figure 2C).

## Genes with bivalent promoters tend to be expressed in a tissue-specific manner

It has been proposed that bivalent promoters poise genes for activation or repression. To investigate whether genes with bivalent promoters were more likely to be differentially expressed upon differentiation, we collected RNA sequencing data for 34 tissues and primary cell lines (Figure 4A). Most of the genes with H3K4me3-only promoters showed a strong expression in all analyzed tissues/cell lines, whereas most genes with H3K27me3-only promoters were not expressed. These findings suggest that the stem cell expression pattern of these two gene categories tend to be maintained also following differentiation. Genes associated with bivalent promoters showed a more dynamic expression pattern. While most of them displayed a low basal expression in hESCs, their expression was variable (from silent to highly expressed) in the different tissues/cells under study, indicating that such genes tend to be expressed or firmly repressed in a tissue-specific manner. We also observed variability between tissues in the overall trend toward expression or repression. For instance, most genes associated with bivalent promoters in hESCs were repressed in hematopoietic cell lines (*e.g.*, CD4 and CD8 naive, CD4 memory cells) and highly expressed in brain tissues (*e.g.*, fetal brain and hippocampus) (Figure 4B). Stritingly, the bivalent promoters of genes that were poorly expressed in all analyzed tissues belonged mainly to cluster 2. Conversely, cluster 1 and 3 bivalent promoters were more often found among highly expressed genes. Genes with cluster 1 promoters were the most highly expressed in most tissues (Figure 4C). This observation suggests that genes associated with PolII-enriched bivalent promoters in hESCs are more prone to be highly expressed upon differentiation.

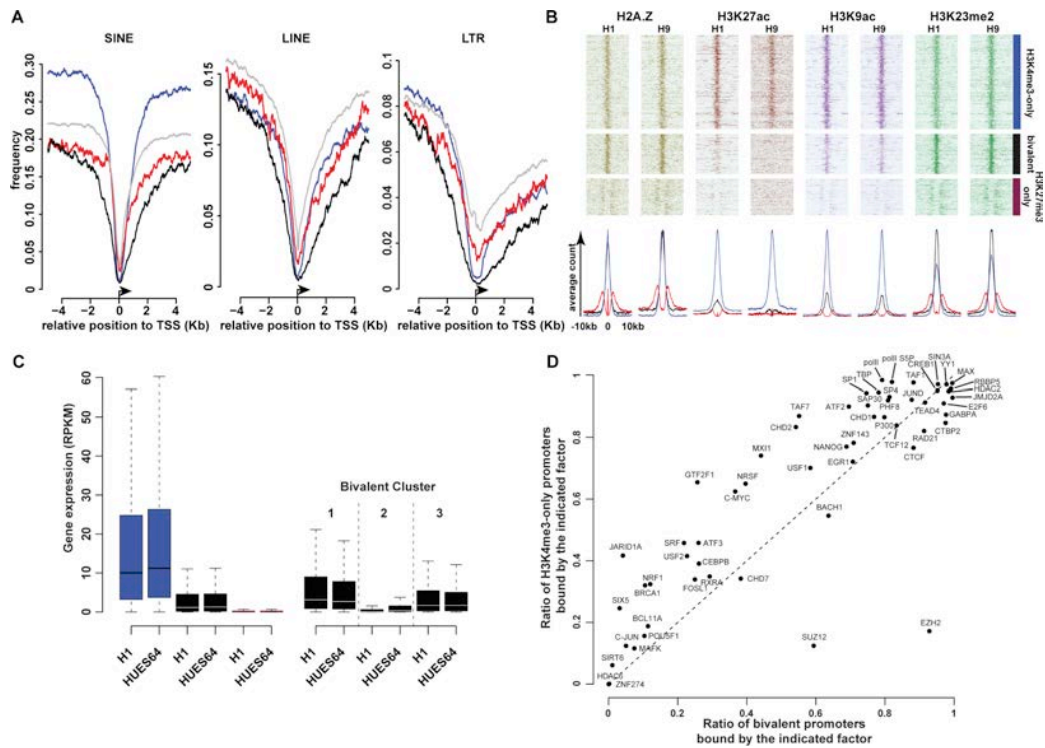

**Figure 2: Main molecular signatures associated with high-confidence H3K4me3-only, bivalent and KH3K27me3-only regions.** (**A**) Frequency of retroelements at all (gray lines); H3K4me3-only (blue lines); bivalent (black lines) and H3K27me3-only (purple lines) promoters (± 4 Kbp from the TSS). (**B**) Occupancy of histone marks at H3K4me3-only; bivalent and H3K27me3-only promoters in H1 and H9 hESC lines. Each line represents one promoter; ChIP-seq read densities for a ± 10Kbp window are centered on the peak signal. The associated average counts are shown in the lower panel. (**C**) Expression level of genes with H3K4me3-only (blue columns), bivalent (black columns) or H3K27me3-only (purple columns) promoters in H1 and HUES64 hESC lines (left histogram). The right histogram shows the expression level of genes with bivalent promoters belonging to cluster 1, 2 and 3, respectively. (**D**) Transcription factor occupancy at bivalent promoters and in H3K4me3-only promoters.

**Integration of the hESC chromatin signature features in customized DNA methylation analyses using Illumina arrays**

To develop a user-friendly approach to integrate these hESC chromatin signature features in customized DNA methylation analyses, we created a text file (Supplementary File S4) ready for uploading onto GenomeStudio, the software dedicated to methylation analyses performed with the HM450K and MethylationEPIC arrays from Illumina. These arrays are widely used in laboratories worldwide and are becoming the gold standard for high-throughput DNA methylation analyses of human specimens. The HM450K array covers 482421 CpG sites and allows interrogating 98.9% of HC bivalent regions (mean coverage, MC: 10.9 probes per region), 88.1% of H3K4me3-only regions (MC: 8.7 probes/region) and 55.7% of H3K27me3-only regions (MC: 2.4 probes/regions). Coverage is higher with the EPIC arrays that allow interrogating 863904 CpG sites, with 99.5% of HC bivalent regions (MC: 12.4 probes/region), 93.3% of H3K4me3-only regions (MC: 10.3 probes /regions) and 73.5% of H3K27me3-only (MC: 3.1 probes/region) (Supplementary Table S7).

After uploading in GenomeStudio, our text file (Supplementary File S4) allows selecting the probes that belong to all or a selected bivalent, H3K4me3-only or H3K27me3-only region from any loaded HM450K or MethylationEPIC dataset. This text file also allows selecting probes belonging to one of the three clusters we defined for bivalent promoters. In addition, functionalities already present in GenomeStudio can be used to filter for probes associated with a variety of genomic features [*e.g.*, CGI (core, shore, shelf), promoters, intergenic…] in order to perform fully customized and detailed analyses (Figure 5A).

By using this text file, we first analyzed the DNA methylation pattern at CGIs relative to their chromatin signature in hESCs. HM450K data for H1 hESCs were downloaded from Encode (see Material and Method section). This array covered a total of 21 177 CGIs analyzed by nearly 146 000 probes. About 36% of these probes were located in H3K4me3-only regions, 26% in bivalent regions and only 1.3% in H3K27me3-only regions. This further confirms that most CGIs are not marked by H3K27me3 alone. The remaining probes (36%) were located in CGIs not marked by any of these three chromatin signatures in hESCs (referred as "none-regions"). In agreement with the unmethylated status of most CGIs, most of these probes were unmethylated (β value < 0.1). Specifically, 90% and 77% of probes located in H3K4me3-only and bivalent regions, respectively, and 38.5% of probes in none-regions displayed a β value < 0.1. Conversely, probes in H3K27me3-only regions were methylated and more than 65% had a β value > 0.7. However, given the low number of probes present in these regions, most of the methylated probes were actually in none-regions (Figure 5B). This suggests that the subset of methylated CGIs in hESCs is depleted in canonical histone modifications (H3K4me3 and H3K27me3). Concerning the genomic features associated with CGIs, unmethylated probes (β value < 0.3) were mainly located in CGIs with promoter features, regardless of the chromatin signature (none, H3K4me3-only or bivalent). Conversely, methylated probes that were mostly in none- and H3K27me3-only regions, to a lesser extent, tended to be present mainly in GCIs located in the gene body (Figure 5C).

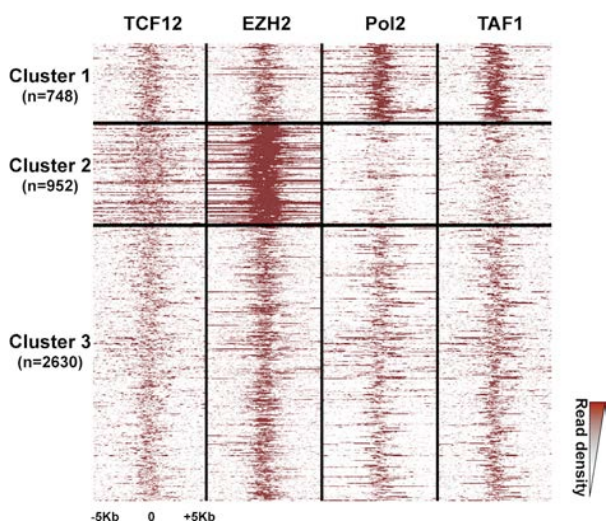

**Figure 3: Three clusters of high-confidence bivalent promoters in human ES cells.** High-confidence bivalent promoters are classified in three clusters based on the extent and density of the EZH2, PolII, TAF1 and TCF-12 signals. Each line represents one promoter; ChIP-seq read densities are shown for a ± 5Kbp window centered on the peak signal.

## Bivalent CGIs associated with genes that are less prone to be expressed upon differentiation are the main target of aberrant hypermethylation in cancer

As a proof of concept, we next evaluated the relationship between hESC chromatin signatures and DNA methylation in cancer. To this aim, we downloaded from the TCGA data portal the HM450K data for eight solid tumor types and their matched normal controls (from 21 matched couples for bladder urothelial carcinoma up to 90 matched couples for breast invasive carcinoma) (Supplementary Table S8). The overall DNA methylation pattern observed in hESCs is maintained in normal tissues with the majority of CGIs marked by bivalency and H3K4me3-only in hESCs remaining unmethylated (Supplementary Figure S6). Compared with normal tissues, in tumor samples we detected DNA methylation changes at CGIs predominantly in bivalent (black) and none-regions (gray), with an overall gain of methylation (increased median value). Conversely H3K4me3-only regions (blue) were generally unaffected (unmethylated in both tumor and control sample) (Figure 6A). In addition, at none-regions, the DNA methylation level distribution tended to be more more widely distributed in tumor samples than in matched controls (bimodal distribution), suggesting that, despite the overall gain of methylation, both gain and loss of methylation occur at these regions in tumor samples (Figure 6A and not shown).

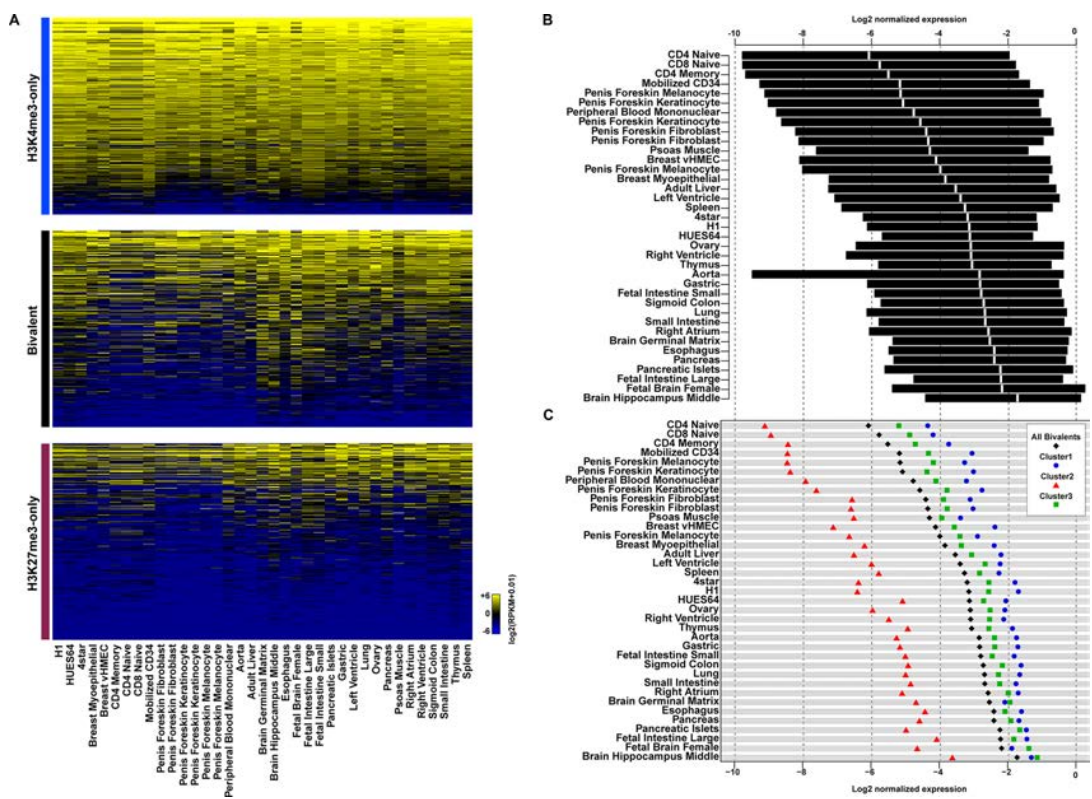

**Figure 4: Genes with bivalent promoters in human ES cells are differentially expressed in committed cell lineages.** (**A**) Expression level in tissues and primary cell lines of genes with H3K4me3-only (upper panel), bivalent (middle panel) and H3K27me3-only (lower panel) promoters in hESCs. Expression level (in RPKM) is denoted by the color scale (blue to yellow: −6 to +6 on a log$_2$ scale). (**B**) Interquartile range of expression in tissues and primary cell lines of genes with a bivalent promoter in hESCs. The median is shown in white. (**C**) Median expression in tissues and primary cell lines of genes with bivalent promoters (all, black diamonds), bivalent promoters belonging to cluster 1 (blue circles), cluster 2 (red triangles) or cluster 3 (green squares). In (B) and (C) gene expression level was normalized to the median of the expression of H3K4me3-only promoter-associated genes from the same tissue.

We then observed that hypermethylated probes (delta β value > 0.25 tumor vs control, FDR < 0.05) were greatly enriched at bivalent regions, irrespectively of the tumor CIMP status. Indeed, in both CIMP-positive and CIMP-negative tumors, most of the hypermethylated probes at CGIs were in bivalent regions, particularly in those belonging to cluster 2 (Figure 6B and 6C). This increased to more than 70% in colon adenocarcinoma (COAD in Figure 6B; Supplementary Table S9) (to be compared with 26% of "bivalent" probes at CGIs on the HM450K array). Conversely, H3K4me3-only regions were under-represented. Although they accounted for a third of all CGI probes on the array, they represented only few percents of the hypermethylated probes in all cancer types (Figure 6B; Supplementary Table S9). Finally and although less prominent than hypermethylation, we also observed hypomethylated probes at CGIs in tumors (delta β value < 0.25 tumor vs control, FDR <0.05), mainly in none-regions.

We next determined the genomic features associated with hypermethylated probes in the eight tumor types (Figure 6D). The distribution of hypermethylated probes was similar in CIMP-positive and CIMP-negative tumors and they were mainly located in promoter regions. This distribution mainly reflects that of all probes in HM450K, indicating that aberrant hypermethylation is equally present at CGIs located in promoter regions and in intergenic regions or gene bodies. The only exception was the H3K4me3-only regions where a relative depletion of hypermethylated probes was observed at promoter regions.

We then collected publicly available RNA sequencing data for healthy liver, lung, breast and colon tissues to determine whether the gene transcriptional status in healthy tissues can influence the gene CGI/promoter methylation status in the corresponding tumor. Overall, gene expression was significantly lower in normal tissues for genes with methylated CGI/promoter in the corresponding tumor, regardless of the CIMP status, compared with gene with unaltered DNA methylation patterns (Figure 6E, Supplementary Figure S7). This suggests that promoters of poorly expressed or repressed genes in healthy tissues are more likely to be aberrantly methylated in tumors.

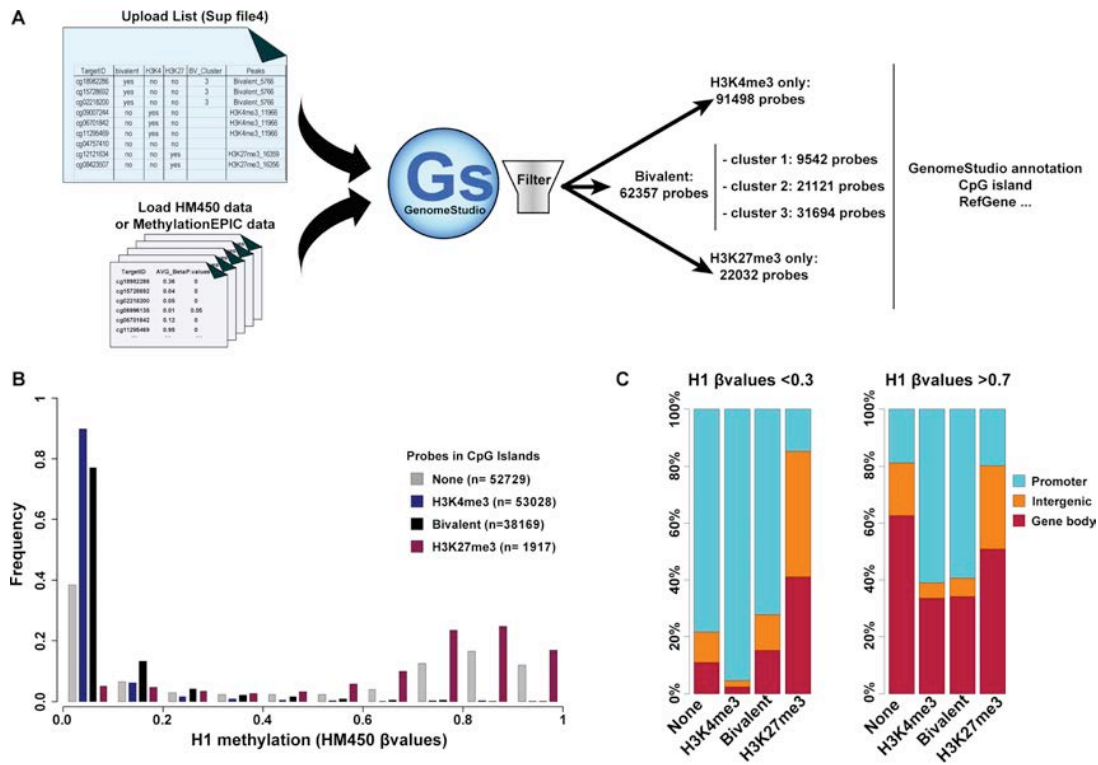

**Figure 5: Integration of the ES cell chromatin signature parameters in DNA methylation analyses.** (**A**) Schematic overview of the approach to integrate hESC chromatin signature parameters in HM450K and MethylationEPIC (Illumina) array-based DNA methylation analyses. (**B**) Methylation status of CGIs in H1 hESCs according to their chromatin signature; n, number of probes per chromatin signature category. (**C**) Genomic features associated with unmethylated (β value < 0.3) and methylated (β value > 0.7) probes in H1 hESC CGIs according to their chromatin signature.

**Aberrant methylation of homeobox gene promoters is a common feature of solid cancers**

Finally, to identify classes of genes that are recurrently affected in cancer, we selected probes that were hypermethylated in at least five of the eight solid tumor types analyzed in this study (4360 probes in CIMP-positive tumors and 97 in CIMP-negative tumors that covered 1596 and 84 promoter regions, respectively) (Figure 7A). In agreement with our previous findings, most of these promoters were in bivalent regions and they belonged mainly to cluster 2 and to a lesser extend to cluster 3 (Figure 7B). Ontology analyses revealed that homeobox genes, including HOX genes, were over-represented among the genes with a recurrently hypermethylated promoter in CIMP-positive tumors ($p$ value < $10^{-40}$) as well as in CIMP-negative tumor types ($p$ value < $10^{-8}$), despite the small number of analyzed genes (Figure 7C).

This suggests that hypermethylation of homeobox gene promoters could constitute a pan-cancer signature.

# DISCUSSION

The currently available lists of genomic regions marked by bivalent chromatin in hESC lines are limited to TSS, are only partially overlapping and are poorly informative and not adapted to determine genome-wide stem cell chromatin signatures that pre-mark CGIs for hypermethylation in cancer cells.

With the objective to facilitate studies on the molecular bases and the impact of aberrant DNA methylation in cancer cells, we integrated publicly available datasets obtained from five independent hESC lines to establish annotated lists of HC bivalent ($n$ = 5 766), H3K4me3-only ($n$ = 11 966) and H3K27me3-only regions ($n$ = 16 361). These numbers are in agreement with a recent study that identified 4 979 bivalent domains at promoter regions in hESCs [24] (4 374 in our study), of which about 81% are also present in our list. The lists we produced include the genomic coordinates and the main features of each region and can be easily used by researchers and clinicians.

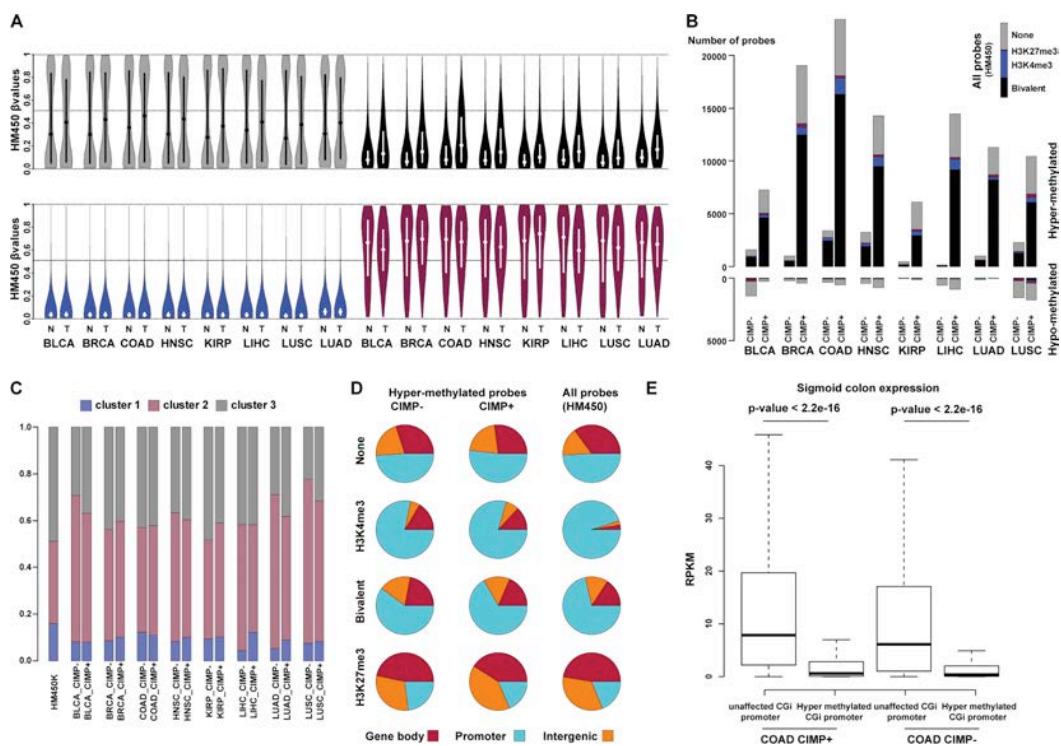

**Figure 6: Bivalent CGIs from cluster 2 are the main target of aberrant hypermethylation in cancer.** (**A**) Violin plots of CGI β value for none (gray), bivalent (black), H3K4me3-only (blue) and H3K27me3-only (purple) regions in eight types of solid tumors (T) and matched normal (N) tissues. The mean β values are plotted on the y-axis, with the median indicated by a dot in the violin. (**B**) Distribution of hyper- and hypo-methylated probes in tumor and matched control tissues according to their chromatin signatures in hESCs (none: gray; bivalent: black; H3K4me3-only: blue; H3K27me3-only: red). For each tumor type, CIMP-positive (+) and CIMP-negative (−) samples were analyzed separately. As a reference, the distribution of all HM450K probes according to their chromatin signatures in hESCs is shown in the upper right corner. (**C**) Relative distribution of hypermethylated probes at bivalent regions (cluster 1 to 3). (**D**) Genomic features associated with hypermethylated probes according to the chromatin signature in hESCs and the CIMP status of the analyzed tumor samples (merged for all tumor types). Distribution on the whole HM450K array (all probes) is shown as reference. (**E**) Genes with aberrantly hypermethylated CGI/promoter in colon adenocarcinoma tend to be repressed in healthy colon tissue. Boxplot representation of the expression levels measured in healthy colon for genes with unaffected or hypermethylated CGI/promoter in CIMP-positive (+) and CIMP-negative (−) colon adenocarcinoma samples ($p$-value: Mann-Whitney test). BLCA: bladder urothelial carcinoma; BRCA: breast invasive carcinoma; COAD: colon adenocarcinoma; HNSC: head-neck squamous cell carcinoma; KIRP: kidney renal papillary cell carcinoma; LIHC: liver hepatocellular carcinoma; LUSC: lung squamous cell carcinoma; LUAD: lung adenocarcinoma.

A number of features, including enrichment in retrotransposons, H3K27me3 read density patterns, associated chromatin signatures and functions of the associated genes, allow discriminating bivalent regions from H3K4me3- and H3K27me3-only regions. This indicates that the detected chromatin bivalency is unlikely to be the result of a mixed hESC population. In addition, the very low expression levels of bivalent region-associated genes in hESCs (10-fold lower compared with H3K4me3-only region-associated genes) and their marked propensity to gain aberrant DNA methylation in cancer cells are two strong evidences against a mixed population.

Moreover, while some permissive histone marks, such as H3K27ac, are only found associated with H3K4me3-only regions, others can be present at both bivalent and H3K4me3-only regions. This include, as previously documented, the histone variant H2AZ [21, 24] and some acetylated histone forms, indicating that, similarly to H3K4me3, a subset of active histone modifications can co-exist with the repressive H3K27me3 mark in a poised configuration. Intriguingly, our analysis also revealed that H3K23me2, a recently identified new heterochromatin mark in *C. elegans* [28], highly marks bivalent regions and, to a lesser extent, H3K4me3-only regions.

Our data further support and extend previous findings [19, 24] showing that CGIs in hESCs are associated with three main chromatin states: H3K4me3-only, bivalent and no mark (*i.e.*, no H3K4me3 and/or H3K27me3). By contrast, H3K27me3-only regions are CGI-depleted. This suggests that different mechanisms are involved in H3K27me3 deposition. In the emerging model indeed, an unmethylated and transcriptionally inactive status at CpG island is sufficient to promote the recruitment of the polycomb responsive complex PRC2 [29–31]. As a result, bivalency is the default signature of transcriptionally inactive CGIs. Recent studies suggest that this model could also apply to non-pluripotent stem cells [32, 33]. Specifically, during somatic development, gain or loss of chromatin bivalency at the key CGIs that regulate imprinting, the so-called imprinting control regions (ICR), contributes to finely tune imprinted gene expression [32]

Beside this CGI-based recruitment, the genomic distribution of H3K27me3-only regions highlights the existence of an alternative mechanism. Indeed, H3K27me3-only regions could also be the secondary result of H3K27me3 spreading from CGIs. Our observation that the H3K27me3 read density at these regions is low and widespread suggests a basal and regional enrichment for this mark, supporting this hypothesis (Figure 1E and S1).

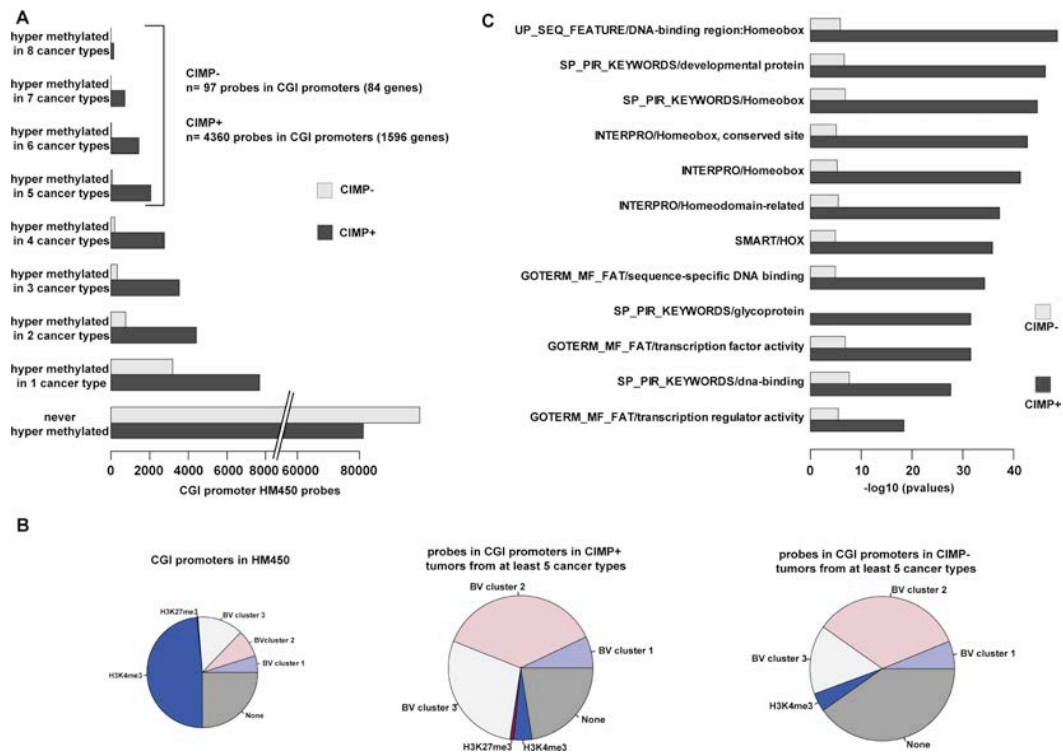

**Figure 7: Hypermethylation of homeobox genes is a pan-cancer signature.** (**A**) Number of hypermethylated probes in eight solid cancer types, classified as CIMP-positive (+) and CIMP-negative (−). The number of hypermethylated probes in five to eight cancer types and the associated number of genes are indicated on the right. (**B**) Distribution of hypermethylated probes in five to eight cancer types (classified as CIMP+ and CIMP− samples) according to the chromatin signature in hESCs. Their distribution in the whole HM450K array is shown as reference (left pie chart). (**C**) Gene ontology terms enrichment for genes with promoter hypermethylated in at least five cancer types (CIMP+ and CIMP− samples).

In agreement with the canonical model in which bivalent promoters in ESCs are poised for activation or repression upon development [9, 26], we observed that unlike H3K4me3- and H3K27me3-only regions, bivalent region-associated genes tend to be expressed in a tissue-specific manner. Interestingly, the overall trend toward gene expression or repression highly varies between cell and tissue types, particularly between hematopoietic cell lines and brain tissues where genes associated with bivalent promoters in hESCs tend to be repressed and strongly expressed, respectively. This probably reflects the variability and complexity of the developmental pathways involved in the commitment and/or maintenance of each cell lineage. Strikingly, the propensity for a gene to be expressed in somatic tissues appears to be dependent on the presence at its promoter of a pre-loaded unproductive transcription complex in hESCs. Specifically, bivalent promoters associated with high PolII and TAF1 occupancy in hESCs (cluster 1 in this study) are more prone to be expressed upon development. Promoters in this cluster could respond more effectively and efficiently to environmental cues and stress conditions.

A key aspect of our work lies in the development of a user-friendly approach to further integrate the stem cell chromatin signature features, collected in our annotated lists, in the widely used Illumina array-based DNA methylation analyses. By this mean, we observed that, as expected [34], the vast majority of CGIs are unmethylated in hESCs and that the subset of methylated CGIs belong almost exclusively to the "none-region" category (*i.e.,* depleted in both H3K4me3 and H3K27me3) and are mainly localized in intergenic and gene body regions. This is in agreement with the observation that these two marks are anti-correlated with DNA methylation at CGIs in mammalian genomes [10, 35–37]. Specifically, as genomic sequences enriched for H3K4me cannot recruit the *de novo* methylation machinery [38–39], this mark is instrumental for protecting CGIs from DNA methylation. Of note, the "none-region" CGIs show a bimodal distribution and those in promoter regions are mainly unmethylated. This observation suggests the existence of an alternative mechanism to protect CGIs from DNA methylation. Further studies should determine whether this protection mechanism relies on H3K4me3-independent transcriptional activity and/or on H3K4me3/H3K27me3-independent poised configuration at these CGI/promoters.

In a proof-of-concept study, conducted using data from eight solid tumor types, we confirmed that aberrant cancer-associated DNA hypermethylation targets mainly CGIs that carry a bivalent signature in hESCs. We also found that aberrant DNA hypermethylation affects specifically CGIs belonging to cluster 2. Consistently, we show that promoter hypermethylation of HOX genes, which are a paradigm of bivalent region-associated genes [9], could constitute a pan-cancer signature. Besides chromatin signatures, our results also support the hypothesis that CGI/promoters of genes repressed in healthy tissues are more prone to DNA methylation in cancer cells, indicating that only a subset of genes with hypermethylated promoters are misregulated in cancer cells (*vs* normal) and contribute to tumor initiation or progression. Our annotated lists can facilitate the identification of these "driver" genes by restricting the possible candidates to hypermethylated genes that are associated with H3K4me3-only and bivalent (cluster 1) CGI/promoter regions in hESCs. We observed that these two categories of genes are expressed in most tested tissues and, therefore, their aberrant methylation is more likely to have a functional impact that could contribute to the tumorigenic process.

Our observations also raise the question of whether the propensity of some CGI/promoters to be aberrantly methylated is a direct consequence of the presence of a bivalent signature in hESCs or the result of that a subset of genes with bivalent promoter in hESCs tend to be transcriptionally silent in most healthy tissue, predominantly those belonging to the cluster 2. Both mechanisms could be involved; however, we favor a model in which the presence of a bivalent signature is a key determinant of the instructive program for aberrant DNA methylation. This is supported, in part, by our finding that bivalent CGI/promoters are the main target of aberrant hypermethylation in both CIMP-positive and CIMP-negative tumors. Specifically, we propose that alterations in the control of the bivalent signature upon differentiation, for instance due to defects in factors involved in bivalency dynamics, could lead to aberrant DNA methylation. Candidate factors include the TET methylcytosine dioxygenases [40] and their catalytic product, the 5hmC that has recently been shown to mark promoter in colon that resist to aberrant DNA methylation in cancer [41]. Histone demethylases are also strong candidate factors. This hypothesis is sustained by the observation that mutations in isocitrate deshydrogenase (IDH) that affect histone demethylation [42] are sufficient to establish a CIMP-positive status in glioma, the main brain tumor type [43]. Similarly, alterations in factors that control only a small subset of bivalent CGIs could lead to a CIMP-negative status. Importantly, in our model, DNA hypermethylation is only one facet of bivalent domain deregulation because this could also lead to DNA methylation-independent defects. Further integrative studies on tumor samples combining DNA methylation, histone modifications and transcriptional analyses are needed to test this hypothesis.

Altogether we provide here a relevant and easy-to-use tool to integrate hESC chromatin signature features in DNA methylation studies. In addition to cancer research, it can be used in a variety of DNA methylation-based customized analyses, for instance the age-associated epigenetic drift.

## MATERIALS AND METHODS

### Creation of high-confidence (HC) lists of bivalent, H3K4me3 and HK27me3 regions in hESCs

Chip-sequencing data for input, H3K4me3 and H3K27me3 immunoprecipitations, aligned on the hg19 genome assembly were obtained from the NIH Roadmap Epigenomics project (http://www.roadmapepigenomics.org/) for five different hESC lines (H1, I3, HUES6, HUES48 and HUES64). Datasets of experiments using the same cell line were combined to reach a minimal sequencing depth of 20 million uniquely mapped reads per experiment (Supplementary Table S2). They were then processed with Macs 1.4.2 using the parameters presented in Supplementary Table S2, as recommended in [44]. Input controls were used for peak detection. For each hESC line, H3K4me3 and H3K27me3 peaks were analyzed with an in-house R scripts. HC bivalent regions were defined as regions where H3K4me3 and H3K27me3 peaks overlapped for at least 1Kbp and in all five hESC lines. By using the same criteria (*i.e.*, more than 1Kbp in size and common to all five hESC lines) and after exclusion of regions defined as bivalent in at least one hESC line, lists of HC H3K4me3-only and H3K27me3-only regions were generated. As the used hESC lines were derived either from female or male embryos, sex chromosomes were excluded from the analysis.

### Annotation of lists of HC regions

All databases and annotations were retrieved for analysis with the hg19 genome assembly. The positions of genes, repeated elements and CGIs were downloaded from UCSC Gencode Comprehensive V19, RepeatMasker and CpG island tracks, respectively. For each gene, promoter regions were defined as TSS±1Kbp. Enhancer positions were determined with the human_permissive_enhancers_phase_1_and_2.bed from the Fantom5 project. HM450K and MethylationEPIC array probe positions were based on Illumina annotations extracted with the GenomeStudio software from the manifest files HumanMethylation450_15017482_v.1.2.bpm and MethylationEPIC_v-1-0_B1.bpm. Genomic features were associated with each HC region using the R Bioconductor package "GenomicRanges".

### Histone modifications and transcription factors

H1 and H9 hESC ChIP-sequencing data for 22 histone marks were obtained from the NIH Roadmap Epigenomics project (Supplementary Table S5). Peaks were called with Macs 1.4.2, using the parameters given in Supplementary Table S5. Histone marks were assigned to an HC region only if an overlapping peak was present in both H1 and H9 cells. Histone mark signals for heatmaps or plots were derived from wig files generated with Macs 1.4.2.

Raw data (SRA) files for 59 transcription factor ChIP-seq experiments using H1 cells were downloaded from Gene Expression Omnibus (GEO) Datasets (Supplementary Table S6). These files were converted with fastq-dump.2.4.1 and aligned to the hg19 genome assembly with bowtie2 (default parameters). Alignments were than processed with Macs 1.4.2 using the parameters listed in Supplementary Table S6. For annotation, peaks with FDR < 0.05 were intersected with HC regions with the R Bioconductor package "GenomicRanges".

To identify HC bivalent region subtypes, K-means clustering was computed for a dozen of candidate protein factors. Relevant partitioning was obtained for four transcription factors: EZH2 (GSM1003524), TAF1 (GSM803450), PolII (GSM803366) and TCF12 (GSM803427). Each HC bivalent region (± 1Kbp window) was defined by values corresponding to the global signal (RPM) for each of these four transcription factors. K-means clustering was then computed using these four values.

### Gene expression data and analysis

RPKM expression matrices for protein-coding genes, non-coding RNAs and ribosomal genes of 57 human samples (tissues and cell lines) were obtained from the Roadmap Epigenomics project (http://egg2.wustl.edu/roadmap/web_portal/). A total of 52 298 genes were analyzed. Gene expression was considered only for genes showing a unique HC chromatin profile at all their marked promoter regions (see Supplementary Figure S3, for example).

To take into account the expression dynamics of bivalent region-associated genes in adult tissues, their expression was normalized to the median expression of H3K4me3-only region-associated genes that was globally stable among tissues.

Functional gene enrichment analyses were conducted with the functional annotation tools of the DAVID 6.7 bioinformatics resources or with the GeneRanker tool of the Genomatix software suite (https://www.genomatix.de/).

### Methylation data and analysis

Methylation data were derived from HM450K Illumina array-based studies. B values for the H1 hESC line were obtained from GEO Datasets under the accession number GSM999379. HM450K methylation data of matched normal and tumor samples were from TCGA (http://cancergenome.nih.gov/) for eight cancer types (Supplementary Table S8). These tumor samples were classified according to their CIMP status provided by [45]. Details on loci defining the CIMP status for each tumor type are given in Supplementary Table S10. Differential methylation analyses were performed using the limma R package, as described in [46]. HM450K probes were considered differentially methylated when FDR < 0.05 and when the β value difference between tumor and matched normal sample was higher than 0.25.

## Abbreviations

BLCA: bladder urothelial carcinoma; BRCA: breast invasive carcinoma; CGI : CpG island; CIMP: CpG island methylator phenotype; COAD: colon adenocarcinoma; GEO: Gene Expression Omnibus  HC: High-confidence; hESC : Human embryonic stem cell; HNSC: head-neck squamous cell carcinoma; Kbp: Kilo base pairs; KIRP:  kidney renal papillary cell carcinoma; LIHC: liver hepatocellular carcinoma; LUAD: lung adenocarcinoma; LUSC: lung squamous cell carcinoma; MC: Mean coverage; NIH: National Institutes of Health; PRC2: Polycomb repressive complex 2; RNA-seq: RNA sequencing; RPKM: Reads per kilobase of transcript per million mapped; TAF1: TATA-box binding protein-associated factor1: TCF-12: Transcription Factor 12; TSS: Transcriptional start site; UCSC: University of California, Santa Cruz.

## CONFLICTS OF INTEREST

None.

## GRANT SUPPORT

The Plan Cancer-INSERM (CS14085CS 'Gliobiv'), the Cancéropole CLARA (Oncostarter « Gliohoxas »), Fonds de dotation Patrick Brou de Lauriére, 'Fondation ARC pour la Recherche sur le Cancer' (EML20120904843); La Ligue contre le Cancer comités Puy de Dôme et Ardèche and 'Conseil régional d'Auvergne' (grants awarded to P.A.).

## Authors' contributions

PA and FC designed the study. FC performed bioinformatics analyses. PA and FC analysed the processed data. FC designed and performed figures and tables with input from PA. PA wrote the manuscript with input from FC. Both authors approved the final manuscript.